\RequirePackage{fix-cm}
\documentclass{iopjournal}

\usepackage{amsmath,amssymb,bm}
\usepackage{booktabs}

\newcommand{\cH}{\mathcal{H}}
\newcommand{\cT}{\mathcal{T}}
\newcommand{\BB}{\mathbb{B}}
\newcommand{\ceff}{c_{\rm eff}}
\newcommand{\rd}{\mathrm{d}}
\newcommand{\ri}{\mathrm{i}}
\newcommand{\re}{\mathrm{e}}
\newcommand{\dNa}{\delta N_a}
\newcommand{\Rc}{\check{R}}
\numberwithin{equation}{section}

\begin{document}

\fancyhead[L]{\small\itshape J. Stat. Mech. Theory Exp.}
\fancyhead[R]{\small A.~Sedrakyan}
\noindent{\large\sffamily Journal of Statistical Mechanics: Theory and Experiment}\par
\medskip
\noindent{\scriptsize\sffamily\bfseries PAPER}\par
\smallskip

\title{Non-compact continuum limit of a\\ $\mathbb{Z}_2$-staggered integrable spin-1 chain:\\ $\ceff=3$ and a black-hole spectrum at frozen level}

\noindent\author{\bfseries A.~Sedrakyan}\par\vspace{-2mm}
\noindent\affil{Alikhanyan National Science Laboratory, Br.\ Alikhanian 2, Yerevan 0036, Armenia}\par
\smallskip
\noindent\par

\begin{abstract}
We apply the staggering construction of Arnaudon, Poghossian, Sedrakyan and Sorba to the spin-1 (Fateev--Zamolodchikov) $R$-matrix, with the shift $\pi/2$ of the spectral parameter for which the spin-$\frac12$ model is known to flow to the $SL(2,\mathbb{R})_k/U(1)$ Euclidean black-hole CFT. For arbitrary spin $S$ we obtain, in closed form in terms of $U_q(sl_2)$ projectors, the local Hamiltonian and a quasi-shift operator $\BB$. For $S=1$ we solve the model by Bethe ansatz and find two critical regimes with non-compact continuum limits. For small anisotropy $\gamma$ the ground state is built from real roots and the black-hole CFT with $k=\pi/\gamma$ and $\ceff=2$ is recovered. For large anisotropy the roots form two seas of 2-strings. We compute the bulk energy and Fermi velocity exactly and show that the sector symmetric under the exchange of the two seas coincides with the homogeneous spin-1 chain, which gives $\ceff=3$ and the compact part of the spectrum. The antisymmetric zero mode produces a continuum. Solving the Bethe equations with string deviations for up to $4096$ sites, we find that the continuous quantum number $s$ obeys three relations---for the energy, for the quasi-shift eigenvalue, $\ln B=4\pi s$, and a quantization condition with reflection phase $\Gamma(\frac12-\ri s)/\Gamma(\frac12+\ri s)$---which are those of the black hole at the $\gamma$-independent level $k=4$, decoupled from a compact boson of $\gamma$-dependent radius. We also report a window around $\gamma=\pi/4$ in which small systems have an exceptional ground state.
\end{abstract}

\section{Introduction}
Integrable models with a staggered arrangement of parameters were introduced in \cite{APSS}, where it was shown that the Yang--Baxter structure survives an alternating shift of the spectral parameter combined with an alternation of the anisotropy; the resulting transfer matrices generate local Hamiltonians on zigzag ladders. The construction was extended to $t$--$J$ ladders \cite{AASSS, TSedrakyan-2001}, to $U_q(gl(N))$ \cite{ASSS}, to the $XYZ$ model \cite{AKMSS} and to multi-leg ladders \cite{ASS}, and the thermodynamics of the staggered spin-$\frac12$ $XXZ$ model was analysed in \cite{MS}; see \cite{KBS} for a recent application of inhomogeneous integrable structures. Closely related chains with alternating shifts or alternating spins appeared in \cite{PZ,FR,dVW92}. It was emphasised already in \cite{APSS} that the construction applies to any solution of the Yang--Baxter equation.

For the six-vertex $R$-matrix and the special shift $\pi/2$ the staggered model is equivalent to the critical antiferromagnetic Potts model \cite{JS06}, and it has become the prototype of a compact lattice model with a \emph{non-compact} continuum limit \cite{IJS08}: the spectrum of conformal weights has a continuous component, visible on the lattice through gaps closing as $1/\ln^2 N$. Ikhlef, Jacobsen and Saleur identified the conformal field theory (CFT) as the $SL(2,\mathbb{R})_k/U(1)$ Euclidean black hole \cite{Witten,DVV,MOS,HPT} with level $k=\pi/\gamma$ \cite{IJS12}. This was confirmed and refined by non-linear integral equations \cite{CI}, by studies of corrections to scaling \cite{FS14}, and by the ODE/IQFT approach \cite{BKKL1,BKKL2}, in which the eigenvalue of a quasi-shift operator provides direct access to the continuous quantum number. Non-compact critical behaviour has since been found in other vertex models \cite{EFS,FM12,VJS14a}, with open boundaries \cite{RJS,FG22}, and for other symmetries \cite{FGNR,KL}.

All these examples are built on fundamental representations. A natural question, directly in the spirit of \cite{APSS}, is what happens when the quantum space carries a higher spin. The homogeneous integrable spin-$S$ chains \cite{ZF,KRS,Takhtajan,Babujian,KR} are critical with central charge $c=3S/(S+1)$ \cite{AGSZ,AM,FYF}; their ground states are seas of $2S$-strings whose deviations from the ideal string positions contribute to the finite-size spectrum at leading order \cite{AD,dVW90,KB,KBP}. For $S=1$ the continuum limit is the $c=\frac32$ superconformal theory of a compact boson and a Majorana fermion. One may therefore expect a staggered spin-1 chain to realise a non-compact theory of a different type.

In this paper we construct and solve this model. Our main results are:
(i) for any spin $S$, a closed expression for the local Hamiltonian $\cH$ and for a quasi-shift operator $\BB$ in terms of $U_q(sl_2)$ projectors [Eqs.~\eqref{eq:H}, \eqref{eq:B}];
(ii) for $S=1$, two regimes of the anisotropy with different root structures. In \emph{regime I} ($0<\gamma<\pi/4$) the Bethe equations differ from those of the spin-$\frac12$ model only in the driving term, and the black-hole CFT with $k=\pi/\gamma$, $\ceff=2$, is recovered;
(iii) in \emph{regime II} ($\pi/4<\gamma<\pi/2$) the ground state consists of two seas of 2-strings. We obtain the bulk energy and velocity in closed form and prove that the sector symmetric under the exchange of the two seas is exactly the homogeneous spin-1 chain on $N/2$ sites; hence $\ceff=3$;
(iv) the antisymmetric zero mode is non-compact. From high-precision solutions of the Bethe equations \emph{including string deviations} we find that the scaling dimensions, the quasi-shift eigenvalue and the quantization of the continuous quantum number all take the black-hole form with the level frozen at $k=4$ for all $\gamma$ in regime II, and with a reflection phase that is blind to the $U(1)$ charges [Eqs.~\eqref{eq:xII}--\eqref{eq:BII}];
(v) exact diagonalization reveals a window around the singular point $\gamma=\pi/4$ in which, for the accessible sizes, the lowest level is an exceptional doublet of a different nature (Sec.~\ref{sec:window}).

The paper is organised as follows. Section~\ref{sec:model} defines the model. Section~\ref{sec:BA} contains the Bethe ansatz and regime I, Sec.~\ref{sec:exact} the exact results for regime II, and Sec.~\ref{sec:num} the numerical solution. Section~\ref{sec:disc} discusses open issues and Sec.~\ref{sec:concl} concludes.

\section{The staggered spin-\texorpdfstring{$S$}{S} model}\label{sec:model}
\subsection{\texorpdfstring{$R$}{R}-matrix and transfer matrix}
Let $q=\re^{\ri\gamma}$ and let $P_j$ ($j=0,\dots,2S$) be the projectors onto the spin-$j$ components of the tensor product of two spin-$S$ representations of $U_q(sl_2)$. The spin-$S$ solution of the Yang--Baxter equation \cite{KRS,Jimbo} is
\begin{equation}\label{eq:R}
\Rc(u)=\sum_{j=0}^{2S}\rho_j(u)\,P_j,\qquad \rho_j(u)=\prod_{k=1}^{j}\sin(k\gamma-u)\prod_{k=j+1}^{2S}\sin(k\gamma+u),
\end{equation}
with $\Rc_{12}(u)\Rc_{23}(u+v)\Rc_{12}(v)=\Rc_{23}(v)\Rc_{12}(u+v)\Rc_{23}(u)$ and $\Rc(0)\propto 1$. With $R=\mathcal{P}\Rc$ ($\mathcal{P}$ the permutation) we define, on a chain of $N=2L$ sites, the $\mathbb{Z}_2$-staggered transfer matrix of \cite{APSS} with shift $\pi/2$,
\begin{equation}\label{eq:t}
t(u)=\mathrm{tr}_0\big[R_{0,2L}(u-\tfrac{\pi}{2})\,R_{0,2L-1}(u)\cdots R_{0,2}(u-\tfrac{\pi}{2})\,R_{0,1}(u)\big],\qquad [t(u),t(v)]=0 .
\end{equation}
The shift $\pi/2$ is special because $\rho_j(\pm\frac{\pi}{2})=(\pm1)^{2S}(-1)^j\prod_{k=1}^{2S}\cos k\gamma$, so that
\begin{equation}\label{eq:Pi}
\Rc(\pm\tfrac{\pi}{2})\;\propto\;\Pi\equiv\sum_{j=0}^{2S}(-1)^jP_j,\qquad \Pi^2=1 .
\end{equation}
The involution $\Pi$ is a $q$-deformation of the permutation, to which it reduces, up to the sign $(-1)^{2S}$, for $\gamma\to0$.

\subsection{Local Hamiltonian and quasi-shift operator}
At $u=0$ and $u=\pi/2$ half of the $R$-matrices in \eqref{eq:t} reduce to permutations and the other half to $\Pi$. A short calculation then gives $t(0)\propto\cT\prod_{k}\Pi_{2k-1,2k}$ and $t(\frac{\pi}{2})\propto\cT\prod_{k}\Pi_{2k,2k+1}$, where $\cT$ is the one-site translation, and
\begin{equation}\label{eq:H}
\cH\equiv\frac{\rd}{\rd u}\ln t(u)\Big|_{u=0}+\frac{\rd}{\rd u}\ln t(u)\Big|_{u=\pi/2}
=\sum_{j=1}^{N}\Big[\Pi_{j,j+1}\,h_{j+1,j+2}\,\Pi_{j,j+1}+\kappa_{j,j+1}\Big],
\end{equation}
\begin{equation}\label{eq:hk}
h=\sum_{j=0}^{2S}\Big(\sum_{k=j+1}^{2S}\cot k\gamma-\sum_{k=1}^{j}\cot k\gamma\Big)P_j,\qquad
\kappa=\sum_{j=0}^{2S}\Big(\sum_{k=1}^{j}\tan k\gamma-\sum_{k=j+1}^{2S}\tan k\gamma\Big)P_j ,
\end{equation}
with periodic boundary conditions. Although $t(u)$ is invariant only under two-site translations, $\cH$ has the same form on every site, $[\cT,\cH]=0$. For $S=\frac12$, Eq.~\eqref{eq:H} reduces, up to normalisation and an additive constant, to the Temperley--Lieb Hamiltonian of \cite{IJS08,IJS12}. For $S=1$ one has $h=(\cot\gamma+\cot2\gamma)(P_0-P_2)+(\cot2\gamma-\cot\gamma)P_1$ and $\kappa=\tan\gamma\,(P_1+P_2-P_0)+\tan2\gamma\,(P_2-P_0-P_1)$.

The ratio of the two shift-like operators defines the \emph{quasi-shift}
\begin{equation}\label{eq:B}
\BB\equiv\prod_{k=1}^{L}\Pi_{2k-1,2k}\prod_{k=1}^{L}\Pi_{2k,2k+1}\;\propto\;t(0)^{-1}\,t(\tfrac{\pi}{2}),\qquad [\BB,\cH]=0,\qquad \cT\,\BB\,\cT^{-1}=\BB^{-1}.
\end{equation}
It generalises the operator that plays a central role for $S=\frac12$ \cite{IJS12,BKKL1}. By the last relation, eigenvalues $B\neq\pm1$ come in pairs $(B,B^{-1})$ of levels degenerate in energy.

The operators $P_j$ are not Hermitian for $|q|=1$, and neither is $\cH$. Its spectrum contains complex-conjugate pairs, but the low-lying levels in the sectors studied below are real, as for $S=\frac12$. $\cH$ commutes with $S^z=\sum_jS^z_j$, and a twist $\varphi$ can be introduced in the usual way through the boundary condition. For $S=1$ the only singular point of $\cH$ in $0<\gamma<\pi/2$ is $\gamma=\pi/4$, where $\cos2\gamma=0$: there $\Rc(\pm\frac{\pi}{2})$ is not invertible and the driving term of the Bethe equations below becomes trivial. (At $\gamma=\pi/3$ the projectors $P_0$ and $P_2$ are separately singular, but $\Pi=1-2P_1$ and $\cH$ remain finite.) We consider $+\cH$ throughout.

Equations \eqref{eq:H}--\eqref{eq:B}, the Yang--Baxter equation and $[t(u),t(v)]=0$ have been verified numerically for $S=\frac12,1,\frac32$ on small chains.

\section{Bethe ansatz for spin 1}\label{sec:BA}
\subsection{Bethe equations, energy and quasi-shift}
The staggered model is an inhomogeneous spin-$S$ vertex model with inhomogeneities $0,\frac{\pi}{2}$, and the algebraic Bethe ansatz applies without change. In the sector $S^z=2SL-M$ the eigenstates are labelled by roots $\lambda_1,\dots,\lambda_M$ solving
\begin{equation}\label{eq:BAE}
\left[\frac{\sinh(2\lambda_j+2\ri S\gamma)}{\sinh(2\lambda_j-2\ri S\gamma)}\right]^{L}=\re^{\ri\varphi}\prod_{k\neq j}\frac{\sinh(\lambda_j-\lambda_k+\ri\gamma)}{\sinh(\lambda_j-\lambda_k-\ri\gamma)},
\end{equation}
and the eigenvalues of $\cH$ and $\BB$ are
\begin{equation}\label{eq:E}
E=Ne_c+\sum_{j=1}^{M}\epsilon_0(\lambda_j),\qquad \epsilon_0(\lambda)=2\ri\big[\coth(2\lambda+2\ri S\gamma)-\coth(2\lambda-2\ri S\gamma)\big],\qquad e_c=-2\sum_{k=1}^{2S}\cot 2k\gamma,
\end{equation}
\begin{equation}\label{eq:Bev}
B^{\pm1}=\prod_{j=1}^{M}\frac{\tanh(\lambda_j+\ri S\gamma)}{\tanh(\lambda_j-\ri S\gamma)} .
\end{equation}
The left-hand side of \eqref{eq:BAE} has period $\ri\pi/2$ in $\lambda_j$, so the Bethe equations are invariant under a global shift $\lambda_j\to\lambda_j+\ri\pi/2$, under which $B\to B^{-1}$: this is the action of $\cT$. From now on $S=1$ and, unless stated otherwise, $L$ is even. We have tested \eqref{eq:BAE}--\eqref{eq:Bev} against exact diagonalization for $N\le12$, extracting the roots of individual eigenvectors from the $T$--$Q$ relation; representative comparisons are given in Table~\ref{tab:ED}.

The low-lying states have their roots on or around the two lines $\mathrm{Im}\,\lambda=\frac{\pi}{4}$ and $\frac{3\pi}{4}$ (mod $\pi$), which are exchanged by $\cT$. We write $\lambda=x+\frac{\ri\pi}{4}$ on line 1 and $\lambda=x+\frac{3\ri\pi}{4}$ on line 2, and use the phase functions
\begin{equation}
\theta_c(x)=2\arctan(\cot c\,\tanh x),\qquad \theta^\times_c(x)=2\arctan(\tan c\,\tanh x),\qquad \Phi_a(x)=\theta^\times_a(2x).
\end{equation}

\begin{table}[t]
\caption{Spin-1 chain: eigenvalues of $\cH$ from exact diagonalization and from the Bethe ansatz. $(N_1,N_2)$ are the numbers of real roots (regime I) or of 2-strings (regime II) on the two lines; states with $N_1\neq N_2$ are doublets. Values of $B$ are from Eq.~\eqref{eq:Bev}; those marked $^*$ were also obtained (to all digits shown) by diagonalizing the operator \eqref{eq:B} in the doublet.}
\centering\small
\begin{tabular}{@{}lllllll@{}}
\toprule
$\gamma$ & $N$ & $S^z$ & roots $(N_1,N_2)$ & $E$ (exact diag.) & $E$ (Bethe ansatz) & $B$ \\
\midrule
0.5 & 12 & 0 & real, $(6,6)$ & $-44.90725175$ & $-44.90725175$ & 1\\
0.5 & 12 & 0 & real, $(7,5)$ & $-44.77922861$ & $-44.77922861$ & -- \\
1.1 & 8 & 0 & 2-strings, $(3,1)$ & $-16.75299356$ & $-16.75299356$ & $464.843616^{\,*}$\\
1.1 & 12 & 0 & 2-strings, $(3,3)$ & $-27.04284330$ & $-27.04284330$ & 1\\
1.1 & 12 & 0 & 2-strings, $(4,2)$ & $-25.47819383$ & $-25.47819383$ & $306.010803^{\,*}$\\
1.1 & 12 & 4 & 2-strings, $(2,2)$ & $-21.39431190$ & $-21.39431190$ & 1\\
1.1 & 12 & 4 & 2-strings, $(3,1)$ & $-20.17717885$ & $-20.17717885$ & $277.931399$\\
0.95 & 12 & 0 & 2-strings, $(4,2)$ & $-38.86553804$ & $-38.86553804$ & $1189.106254$\\
1.3 & 12 & 0 & 2-strings, $(4,2)$ & $-28.26131034$ & $-28.26131034$ & $112.841133$\\
\bottomrule
\end{tabular}
\label{tab:ED}
\end{table}

\subsection{Regime I: real roots and the black hole at \texorpdfstring{$k=\pi/\gamma$}{k=pi/gamma}}
\begin{figure}[t]
\centering
\includegraphics[width=0.86\textwidth]{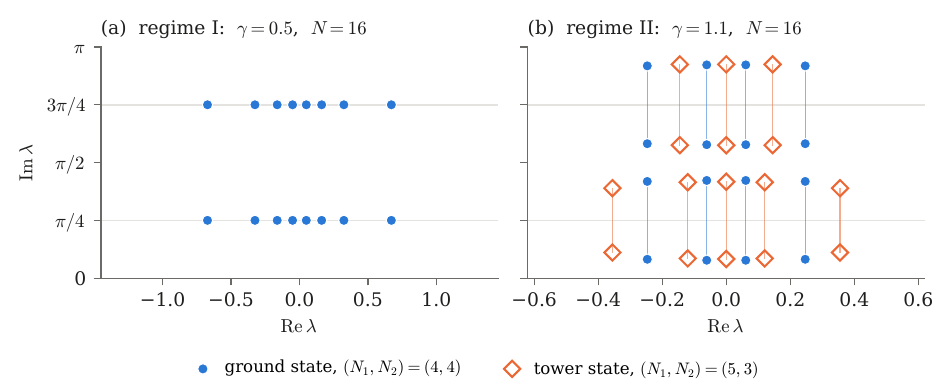}
\caption{Bethe roots in the strip $0\le\mathrm{Im}\,\lambda<\pi$. (a) Regime I: real parts on the lines $\mathrm{Im}\,\lambda=\pi/4,\,3\pi/4$. (b) Regime II: 2-strings centred on the two lines (members joined by thin lines). The tower state has one string moved from line 2 to line 1.}
\label{fig:roots}
\end{figure}

For real $x$ the bare energy on the two lines is $\epsilon_0=-\Phi'_{2\gamma}(x)=-4\sin4\gamma/(\cosh4x+\cos4\gamma)$, which is negative for $\gamma<\pi/4$. For small $\gamma$ the ground state has $L$ real roots on each line [Fig.~\ref{fig:roots}(a)], and Eq.~\eqref{eq:BAE} becomes
\begin{equation}\label{eq:BAE1}
L\,\Phi_{2\gamma}(x_j)=2\pi I_j-\sum_{k\in\text{same}}\theta_\gamma(x_j-x_k)+\sum_{k\in\text{other}}\theta^\times_\gamma(x_j-x_k).
\end{equation}
The right-hand side is \emph{identical} to that of the spin-$\frac12$ model \cite{IJS08}; only the driving term has changed ($\Phi_{2\gamma}$ instead of $\Phi_\gamma$). Since the dressed charges depend only on the kernels, the operator content is that of the spin-$\frac12$ model, i.e.\ the black-hole CFT with $k=\pi/\gamma$ and $\ceff=2$. The driving term merely changes the root density, $\hat\rho(\omega)=\cosh(\gamma\omega/2)/\cosh(\eta\omega/4)$ with $\eta\equiv\pi-2\gamma$, the bulk energy and the non-universal scale $N_0$; the velocity is $v=4\pi/\eta$. We have checked this numerically, solving \eqref{eq:BAE1} for $N\le2048$. We define the effective central charge and the scaled gaps $x$ through \cite{BCNA}
\begin{equation}\label{eq:fss}
E_0=Ne_\infty-\frac{\pi v\,\ceff}{6N}+o(N^{-1}),\qquad E-E_0=\frac{2\pi v}{N}\,x .
\end{equation}
With $n=S^z$, and with $\tilde m=N_1-N_2$ the difference of the numbers of roots on the two lines, we find
\begin{equation}\label{eq:I}
x=\frac{n^2}{2k}+\frac{2s^2}{k-2},\qquad \frac{\pi}{2}\,\tilde m=2s\ln\frac{N}{N_0}+2\arg\Gamma\Big(\frac{1+|n|}{2}-\ri s\Big),\qquad \ln B=\frac{8\pi s}{k-2},\qquad k=\frac{\pi}{\gamma},
\end{equation}
with a single constant $N_0$ for all states (e.g.\ $-\ln N_0=0.8463$ at $\gamma=0.5$ for the seven states with $n\le3$, $\tilde m\le4$ examined), and $\ceff=2.00000(1)$. The second relation is the black-hole quantization condition \cite{MOS,HPT,IJS12}. The third is the spin-1 counterpart of the relation between the quasi-shift and $s$ known for $S=\frac12$ \cite{IJS12,BKKL1}, where we find $\ln B=4\pi s/(k-2)$ with the same normalisation \eqref{eq:Bev}; in general the right-hand side carries a factor $2S$.

At $\gamma=\pi/4$ the driving term of \eqref{eq:BAE} is trivial and for $\gamma>\pi/4$ real roots on the two lines have positive energy: the ground state must reorganise.

\subsection{Regime II: two seas of 2-strings}
For $\pi/4<\gamma<\pi/2$ we find that the roots of the ground state and of the low-lying states with $S^z\equiv0$ (mod 4) form 2-strings centred on the two lines [Fig.~\ref{fig:roots}(b)],
\begin{equation}\label{eq:strings}
x_j^{\pm}=\xi_j\pm\ri\big(\tfrac{\gamma}{2}+\delta_j\big),\qquad \xi_j\in\mathbb{R},
\end{equation}
with $N_1$ strings on line 1 and $N_2$ on line 2, $S^z=2(L-N_1-N_2)\equiv2m'$. The deviations $\delta_j$ are small in the bulk of the seas but of order one at their edges. Neglecting them (Bethe--Takahashi approximation, BT) and multiplying the two equations of a string one obtains
\begin{equation}\label{eq:BT}
L\big[\Phi_{3\gamma}+\Phi_{\gamma}\big](\xi_j)=2\pi I_j-\sum_{k\in\text{same}}\big[2\theta_\gamma+\theta_{2\gamma}\big](\xi_j-\xi_k)+\sum_{k\in\text{other}}\big[2\theta^\times_\gamma+\theta^\times_{2\gamma}\big](\xi_j-\xi_k),
\end{equation}
while the energy of a string is $-[\Phi'_{3\gamma}+\Phi'_\gamma](\xi_j)$. The exact equations for $(\xi_j,\delta_j)$ are given in Appendix~\ref{app}. Section~\ref{sec:exact} collects what can be derived analytically; Sec.~\ref{sec:num} reports the solution of the exact equations.

\section{Regime II: exact results}\label{sec:exact}
\subsection{Bulk properties}
Let $\rho_{1,2}(\xi)$ be the densities of string centres per unit of $L$ and $\rho_\pm=\rho_1\pm\rho_2$. In the absence of holes the continuum limit of \eqref{eq:BT} reads, in Fourier space [$\hat f(\omega)=\int\rd\xi\,\re^{\ri\omega\xi}f(\xi)$],
\begin{gather}\label{eq:symbols}
\mathsf{S}(\omega)\,\hat\rho_+(\omega)=2\hat d(\omega),\qquad \mathsf{A}(\omega)\,\hat\rho_-(\omega)=0,\\
\mathsf{S}=\frac{4\cosh^2G\,\sinh(P-2G)}{\sinh P},\qquad
\mathsf{A}=\frac{4\sinh^2G\,\cosh(P-2G)}{\cosh P},\qquad
\hat d=\frac{2\cosh G\,\sinh(P-2G)}{\sinh P},\nonumber
\end{gather}
where $G=\eta\omega/4$, $P=\pi\omega/4$ and $\eta=\pi-2\gamma\in(0,\frac{\pi}{2})$. The ground state has $\rho_1=\rho_2=\rho$ with $\hat\rho=1/(2\cosh G)$, i.e.
\begin{equation}\label{eq:bulk}
\rho(\xi)=\frac{1}{\eta\cosh(2\pi\xi/\eta)},\qquad N_1=N_2=\frac{L}{2},\qquad e_\infty=-\frac{2}{\sin2\eta},\qquad v=\frac{4\pi}{\eta},
\end{equation}
where $e_\infty$ is the energy per site and $v$ the velocity of the gapless excitations, obtained from the dressed energy $\varepsilon(\xi)=-2\pi\rho(\xi)$. Because the deviations $\delta_j$ vanish exponentially in the bulk of the seas, Eq.~\eqref{eq:bulk} is exact; it is confirmed below to ten digits.

\subsection{The symmetric sector is the homogeneous spin-1 chain}
Consider a solution of \eqref{eq:BAE} that is invariant under $\lambda\to\lambda+\ri\pi/2$: the two lines carry identical sets $\{x_j\}$, and $B=1$. For such a set the product over the roots of both lines combines pairwise, $\sinh(y+\ri\gamma)\cosh(y+\ri\gamma)=\frac12\sinh(2y+2\ri\gamma)$. In the variables $w_j=2x_j-\frac{\ri\pi}{2}$, and using $2\gamma=\pi-\eta$, Eq.~\eqref{eq:BAE} becomes
\begin{equation}\label{eq:FZ}
\left[\frac{\sinh(w_j+\ri\eta)}{\sinh(w_j-\ri\eta)}\right]^{L}=\re^{-\ri\varphi}\prod_{k\neq j}\frac{\sinh(w_j-w_k+\ri\eta)}{\sinh(w_j-w_k-\ri\eta)},
\end{equation}
which are the Bethe equations of the \emph{homogeneous} spin-1 chain \cite{ZF,KR} with anisotropy $\eta$ on $L$ sites, and $E-Ne_c=2\sum_j[-4\sin2\eta/(\cosh2w_j-\cos2\eta)]$ is twice its energy. The 2-strings \eqref{eq:strings} map onto its standard 2-strings, and the window $\pi/4<\gamma<\pi/2$ onto the full critical regime $0<\eta<\pi/2$ of that chain. The argument holds for any $S$, with $[\sinh(w+\ri S\eta)/\sinh(w-\ri S\eta)]^L$ on the left-hand side.

The symmetric sector of the staggered chain on $N=2L$ sites is therefore the spectrum of the homogeneous chain on $L$ sites, with all energies doubled. With $v=2v_{\rm hom}$ and the definitions \eqref{eq:fss}, this implies that the central charge and the scaling dimensions are twice those of the homogeneous chain. The ground state lies in this sector, hence
\begin{equation}\label{eq:c3}
\ceff=2\times\frac32=3\qquad\Big(\text{for general }S\text{ and a symmetric ground state of }2S\text{-strings}:\ \ceff=\frac{6S}{S+1}\Big),
\end{equation}
and the Gaussian part of the spectrum of the spin-1 chain \cite{AM,FYF} yields, for the 2-string states,
\begin{equation}\label{eq:compact}
x_{\rm c}=m'^2X+\frac{e^2}{4X},\qquad X=\frac{\pi-2\eta}{2\pi}=\frac{4\gamma-\pi}{2\pi},
\end{equation}
where $e$ is the shift of the string quantum numbers $I_j$ from the symmetric distribution (Sec.~\ref{sec:num}). The contribution of the string deviations to the central charge, known to raise $c=1$ to $\frac32$ in the homogeneous chain \cite{dVW90,KB,KBP}, is thus inherited; in the BT approximation one would find $\ceff=2$. The full spectrum of the $c=\frac32$ theory, including its Ising-twisted sectors, is embedded in the same way. For example, the lowest state with $S^z=2$ ($m'=1$) at $\varphi=0$ is symmetric and consists of 2-strings plus one real root with $\mathrm{Im}\,\lambda=0$ and one with $\mathrm{Im}\,\lambda=\frac{\pi}{2}$; it maps onto the lowest $S^z=1$ state of the spin-1 chain, whence $x=X+\frac14$ \cite{AM} ($=0.450$ at $\gamma=1.1$; exact diagonalization gives $0.504$ and $0.479$ for $N=8$ and $12$).

\subsection{The antisymmetric zero mode}
The symbol $\mathsf{A}(\omega)$ of the antisymmetric channel has a \emph{double zero} at $\omega=0$, $\mathsf{A}\simeq\kappa\,\omega^2$ with $\kappa=\eta^2/4$. Equivalently, the number of vacancies for strings on line 1 minus $N_1$ depends only on $N_1+N_2$: moving a string from one line to the other creates no hole. Imbalanced configurations $(N_1,N_2)=(\frac{L}{2}-\frac{m'}{2}+\frac{\dNa}{2},\frac{L}{2}-\frac{m'}{2}-\frac{\dNa}{2})$ therefore yield a tower of states above each level of the symmetric sector, with $B\neq1$. A standard Wiener--Hopf analysis \cite{dVW85,IJS08} of the dressed charge of a channel whose symbol has a double zero gives $Z_a\simeq\ln(N/N_0)/(\alpha\sqrt{\kappa})$, where $\alpha=2\pi/\eta$ is the decay rate of $\rho$, and hence
\begin{equation}\label{eq:WH}
x_{\rm BT}\simeq\frac{\dNa^{\,2}}{8Z_a^2}=\frac{\kappa\alpha^2\,\dNa^{\,2}}{8\ln^2(N/N_0)}=\frac{\pi^2\,\dNa^{\,2}}{8\ln^2(N/N_0)} .
\end{equation}
The gaps close logarithmically---the continuum limit is non-compact---and, because $\kappa\alpha^2=\pi^2$, the amplitude does not depend on $\gamma$. This is to be contrasted with $S=\frac12$, where the same formula gives $\kappa\alpha^2=\pi^2\gamma/(\pi-2\gamma)$ and reproduces the level $k-2=(\pi-2\gamma)/\gamma$ of \cite{IJS08,IJS12}. However, Eq.~\eqref{eq:WH} is not exact for $S=1$: the amplitude is controlled by the edges of the seas, where the string deviations are not small. We shall see that the exact amplitude is twice as large.

\section{Regime II: solution of the exact Bethe equations}\label{sec:num}
\subsection{Method and selection rule}
We solve the exact equations of Appendix~\ref{app} for the $2(N_1+N_2)$ real unknowns $(\xi_j,\ln|\sin2\delta_j|)$ by a damped Newton iteration with analytic Jacobian, starting from the BT solution. Residuals below $10^{-11}$ are reached for up to $N=4096$ sites ($2048$ strings), and the solutions reproduce the exact-diagonalization data of Table~\ref{tab:ED}. For each state we evaluate the energy \eqref{eq:E} and the quasi-shift eigenvalue \eqref{eq:Bev}, and the corresponding BT quantities for comparison.

A point requires care. The equation for the centre of a string is the product of the equations of its two roots and fixes them only up to a common sign. We find that the solutions with ``narrow'' strings ($\delta_j<0$) and quantum numbers $I_j\in\{-\frac{N_a-1}{2},\dots,\frac{N_a-1}{2}\}+e$ on line $a$ solve \eqref{eq:BAE} with twist
\begin{equation}\label{eq:selection}
\varphi=\pi\,(m'+e)\quad(\mathrm{mod}\ 2\pi),
\end{equation}
which we have verified at the level of the individual roots for every solution reported here. For the periodic chain the 2-string states thus have $m'+e$ even: $S^z\equiv0$ (mod 4) for the symmetric quantum numbers, while the states with $m'$ odd and $e=0$ belong to the chain with twist $\pi$. Non-integer $e$ interpolates continuously.

\subsection{Results}
Table~\ref{tab:num} shows data for $\gamma=1.1$; the pattern is the same at $\gamma=0.95$ and $1.3$.

\emph{Central charge and compact dimensions.} The ground-state energy gives $\ceff=3.00001$ at $N=2048$ ($3.00000$ at $\gamma=0.95$; $3.00010$ at $N=4096$ for $\gamma=1.3$), which confirms $e_\infty$, $v$ and \eqref{eq:c3}; the BT energies give $\ceff=2$ instead. The states with $\dNa=0$ reproduce \eqref{eq:compact} to five digits, both for removed strings ($m'=2$) and for twisted quantum numbers ($e=\frac14,\frac12$).

\begin{figure}[t]
\centering
\includegraphics[width=0.92\textwidth]{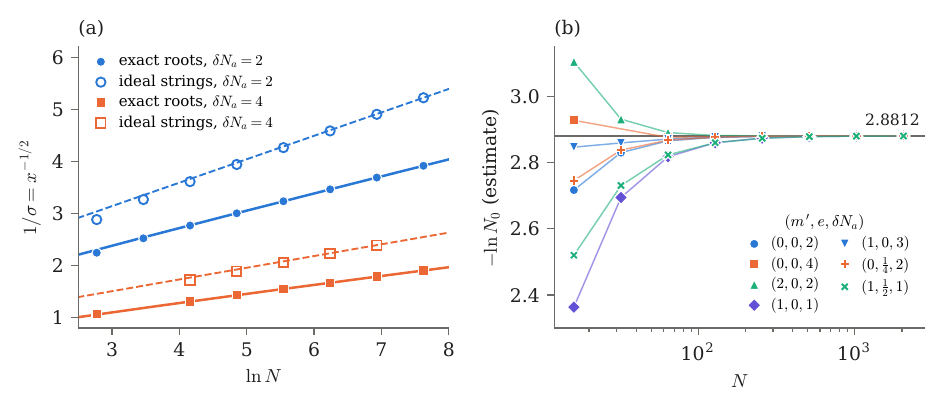}
\caption{Regime II, $\gamma=1.1$. (a) Inverse of $s=\sqrt{x}$ for the tower states $\dNa=2,4$ (with $m'=e=0$). Full symbols: exact roots; solid lines: Eq.~\eqref{eq:qII} with $-\ln N_0=2.8812$. Open symbols: ideal strings; dashed lines have the slope $2\sqrt2/(\pi\dNa)$ of Eq.~\eqref{eq:WH}. (b) Estimates of $-\ln N_0$ obtained from \eqref{eq:xII}, \eqref{eq:qII} for seven states.}
\label{fig:tower}
\end{figure}

\emph{Continuous quantum number.} For $\dNa\neq0$ we define $x_{\rm nc}=x-x_{\rm c}$. All our data are described by
\begin{align}
x&=m'^2X+\frac{e^2}{4X}+s^2, \label{eq:xII}\\
\pi\,|\dNa|&=2s\ln\frac{N}{N_0}+2\arg\Gamma\big(\tfrac12-\ri s\big), \label{eq:qII}\\
\ln B&=4\pi s, \label{eq:BII}
\end{align}
with a non-universal constant $N_0(\gamma)$ that is the same for all states. Equation~\eqref{eq:BII} is satisfied by $s=\sqrt{x_{\rm nc}}$ to $10^{-5}$ at $N=2048$ (Table~\ref{tab:num}). Equation~\eqref{eq:qII} is tested by solving it for $\ln N_0$ state by state: the estimates from seven states with $\dNa=1,\dots,4$ and different $(m',e)$ converge to the same value $-\ln N_0=2.8812$ [Fig.~\ref{fig:tower}(b)], and similarly $-\ln N_0=1.6803$ at $\gamma=0.95$. At $N=2048$ these states have $s=0.126$, $0.255$, $0.388$, $0.525$ for $\dNa=1,2,3,4$, and states with equal $\dNa$ but different $(m',e)$ have the same $s$ to four digits. Since $\arg\Gamma(\frac12-\ri s)/s$ varies by $0.45$ over this range, the agreement to $10^{-4}$ is a stringent test of the functional form of the reflection phase and of its independence of $m'$ and $e$. Figure~\ref{fig:tower}(a) shows the tower itself: $1/s$ grows linearly with $\ln N$ with slope $2/(\pi|\dNa|)$, whereas the BT data have a slope larger by $\sqrt2$, i.e.\ $x_{\rm BT}\simeq x_{\rm nc}/2$, in agreement with \eqref{eq:WH}.

\emph{Corrections to scaling.} For $\gamma=0.95$ and $1.1$ all corrections decay approximately as $N^{-2}$. Closer to $\gamma=\pi/2$ they are governed by the exponent $a=2\eta/(\pi-2\eta)<1$: for $\gamma\ge1.2$ we observe $\ceff-3\propto N^{-2a}$ and corrections $\propto N^{-a}$ to \eqref{eq:xII}--\eqref{eq:BII}. This corresponds to a perturbation of dimension $2+a=1+\pi/(\pi-2\eta)$ in each spin-1 copy, which becomes marginal for $\eta\to0$ and explains the slow convergence near $\gamma=\pi/2$ (at $\gamma=1.3$, $a=0.53$). At $\gamma=1.3$ the estimates of $-\ln N_0$ therefore drift slowly ($4.4139$, $4.4322$ at $N=2048$, $4096$ for $\dNa=2$), but the leading correction is absent for $e=\frac12$, where they are stable ($4.4832$, $4.4837$, $4.4834$ for $N=512$, $1024$, $2048$), and it is common to states with different $\dNa$: using $s=\ln B/4\pi$ in \eqref{eq:qII}, the states $\dNa=2$ and $4$ give the same value, $4.4652$, at $N=2048$, although their $s$ differ by a factor two.

\begin{table}[t]
\caption{Regime II, $\gamma=1.1$ ($X=0.200282$, $v=13.3459$, $e_\infty=-2.10171883$). States are labelled by $(m',e,\dNa)$; the twist follows from \eqref{eq:selection}. The last line gives the predictions $4X$, $1/16X$ and the common limit of the last five columns.}
\centering\small\setlength{\tabcolsep}{4.2pt}
\begin{tabular}{@{}rcccccccccc@{}}
\toprule
 & & & $x$ & $x$ & $\ln B/4\pi\sqrt{x}$ & \multicolumn{5}{c}{$-\ln N_0$ from Eq.~\eqref{eq:qII} for $(m',e,\dNa)=$}\\
$N$ & $\ceff$ & $\ceff^{\rm BT}$ & $(2,0,0)$ & $(0,\frac12,0)$ & $(0,0,2)$ & $(0,0,2)$ & $(0,0,4)$ & $(2,0,2)$ & $(1,0,1)$ & $(0,\frac14,2)$ \\
\midrule
16 & 3.07239 & 2.0507 & 0.80473 & 0.31945 & 0.97795 & 2.7173 & 2.9284 & 3.1054 & 2.3633 & 2.7461 \\
64 & 3.00469 & 2.0033 & 0.80125 & 0.31254 & 0.99840 & 2.8662 & 2.8766 & 2.8911 & 2.8181 & 2.8682 \\
256 & 3.00031 & 2.0002 & 0.80113 & 0.31209 & 0.99989 & 2.8800 & 2.8806 & 2.8809 & 2.8745 & 2.8801 \\
1024 & 3.00002 & 2.0000 & 0.80113 & 0.31206 & 0.99999 & 2.8811 & 2.8812 & 2.8811 & 2.8805 & 2.8812 \\
2048 & 3.00001 & 2.0000 & 0.80113 & 0.31206 & 1.00000 & 2.8812 & 2.8812 & 2.8812 & 2.8810 & 2.8812 \\
\midrule
exact & 3 & (2) & 0.80113 & 0.31206 & 1 & \multicolumn{5}{c}{$2.8812$ (common value)} \\
\bottomrule
\end{tabular}
\label{tab:num}
\end{table}

\subsection{The level}
\begin{figure}[t]
\centering
\includegraphics[width=0.9\textwidth]{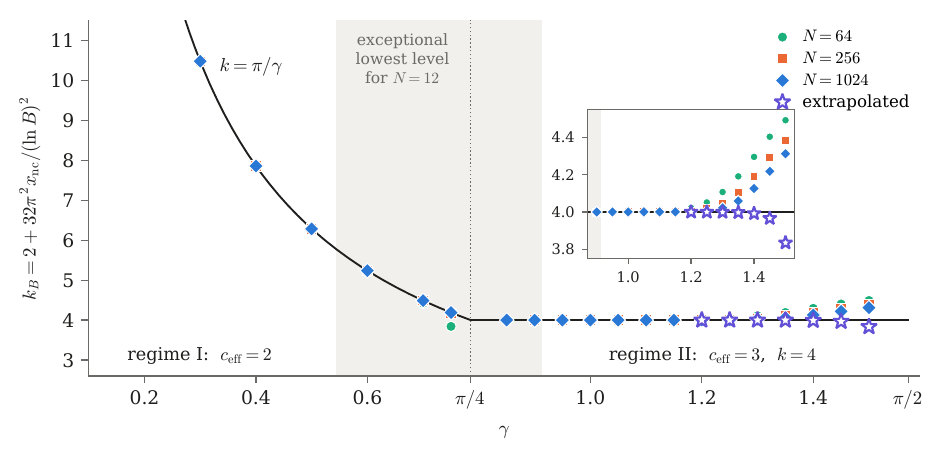}
\caption{Level estimator \eqref{eq:kB} for the lowest tower state of the spin-1 chain ($\tilde m=2$ in regime I, $\dNa=2$ in regime II). Lines: $\pi/\gamma$ and $4$. Stars: extrapolation of the $N=256,1024$ data with a correction $\propto N^{-a}$. Inset: enlargement of regime II. In the shaded window the string states are not the lowest levels for $N=12$ (Sec.~\ref{sec:window}).}
\label{fig:k}
\end{figure}

Equations \eqref{eq:xII}--\eqref{eq:BII} are precisely Eqs.~\eqref{eq:I} of the black-hole CFT, restricted to vanishing black-hole charge $n=0$, with $\tilde m=2\dNa$ (still the difference of the numbers of roots on the two lines) and
\begin{equation}
k=4\qquad\text{for all}\quad \tfrac{\pi}{4}<\gamma<\tfrac{\pi}{2}.
\end{equation}
Three independent relations fix this value: the dimension $2s^2/(k-2)=s^2$, the quasi-shift $8\pi s/(k-2)=4\pi s$, and the fact that the same $s$ is the argument of the reflection phase. A parameter-free estimator of the level, which uses only the scaled gap and the quasi-shift eigenvalue of a single state, is
\begin{equation}\label{eq:kB}
k_B\equiv2+\frac{32\pi^2\,x_{\rm nc}}{(\ln B)^2}.
\end{equation}
Figure~\ref{fig:k} shows $k_B$ for the lowest tower state as a function of $\gamma$ in both regimes. It converges to $\pi/\gamma$ in regime I (e.g.\ $k_B=6.28315$ at $\gamma=0.5$, $N=1024$) and to $4$ in regime II ($|k_B-4|<2\times10^{-4}$ for $0.9\le\gamma\le1.15$ at $N=1024$); close to $\gamma=\pi/2$ the convergence is slow, and an extrapolation with the exponent $a$ given above restores $k=4$ within $1\%$ up to $\gamma=1.45$ (and $4\%$ at $\gamma=1.5$).

The $U(1)$ charges enter very differently in the two regimes. In regime I, $S^z=n$ is the charge of the black hole: it shifts the argument of the reflection phase in \eqref{eq:I}, and its conjugate radius is tied to the level. In regime II, $S^z=2m'$ and the twist are carried by a compact boson whose radius $X$ varies continuously with $\gamma$ and which does not enter \eqref{eq:qII}. Within the 2-string sector the continuum limit of regime II thus has the structure of the neutral sector of $SL(2,\mathbb{R})_4/U(1)$ times a decoupled Gaussian model, with the charge lattice $(m',e)$, $m'+e$ even, at $\varphi=0$. Table~\ref{tab:sum} summarises the two regimes.

\begin{table}[t]
\caption{The two regimes of the staggered spin-1 chain (string sectors; zero twist in regime I). In both, $v=4\pi/(\pi-2\gamma)$ and the continuous quantum number is quantized by $\frac{\pi}{2}\tilde m=2s\ln(N/N_0)+\frac12\delta(s)$, with $\tilde m$ the difference of the numbers of roots on the two lines.}
\centering\small
\begin{tabular}{@{}lll@{}}
\toprule
 & regime I \ ($0<\gamma<\pi/4$) & regime II \ ($\pi/4<\gamma<\pi/2$)\\
\midrule
roots of the ground state & real, on two lines & 2-strings centred on two lines\\
$\ceff$ & 2 & 3\\
$U(1)$ part of $x$ & $n^2/2k$, \ $n=S^z$ & $m'^2X+e^2/4X$, \ $m'=S^z/2$\\
continuous part of $x$ & $2s^2/(k-2)$ & $s^2$\\
quasi-shift, $\ln B$ & $8\pi s/(k-2)$ & $4\pi s$\\
reflection phase $\re^{\ri\delta(s)}$ & $\Gamma(\frac{1+|n|}{2}-\ri s)^2/\Gamma(\frac{1+|n|}{2}+\ri s)^2$ & $\Gamma(\frac12-\ri s)^2/\Gamma(\frac12+\ri s)^2$\\
level & $k=\pi/\gamma$ & $k=4$\\
\bottomrule
\end{tabular}
\label{tab:sum}
\end{table}

\section{Discussion}\label{sec:disc}
\subsection{The neighbourhood of \texorpdfstring{$\gamma=\pi/4$}{gamma=pi/4}}\label{sec:window}
The bare energy $\epsilon_0(\lambda)$ has poles at $\lambda=\pm\ri\gamma$ (mod $\ri\pi/2$), which approach the root lines as $\gamma\to\pi/4$. Exact diagonalization shows that this has consequences for small systems: in a window $\gamma^*_{\rm I}(N)<\gamma<\gamma^*_{\rm II}(N)$ containing $\pi/4$ the lowest level of $\cH$ is neither the real-root nor the 2-string state but an exceptional doublet ($B\ne1$) with no roots on the two lines. Its roots are real, with $\mathrm{Im}\,\lambda=0$ (or $\frac{\pi}{2}$ for the partner), or cluster near the poles $\pm\ri\gamma$, e.g.\ $\lambda=\pm0.035\pm0.913\,\ri$ for $\gamma=0.9$, $N=8$, where they collect a large negative energy. The edges of the window are
\begin{equation}
\big(\gamma^*_{\rm I},\gamma^*_{\rm II}\big)=(0.398,\,0.963),\ (0.517,\,0.924),\ (0.544,\,0.913)\qquad\text{for } N=4,\,8,\,12,
\end{equation}
so that the string states are the ground states for $\gamma\le0.5$ and $\gamma\ge0.95$ at $N=8,12$. The window shrinks as $N$ grows, but three sizes do not allow us to decide whether it closes onto the singular point $\gamma=\pi/4$ or whether a third regime of finite width survives in the thermodynamic limit. No such states occur for $S=\frac12$, where we find the real-root state to be the ground state up to $\gamma=1.55$ ($N\le16$). All results of Secs.~\ref{sec:BA}--\ref{sec:num} refer to the string sectors, which are well defined throughout $0<\gamma<\pi/4$ and $\pi/4<\gamma<\pi/2$; close to $\pi/4^+$ we find imbalanced 2-string solutions only beyond a size that grows as $\gamma\to\pi/4$ ($N\gtrsim500$ at $\gamma=0.85$). The nature of the exceptional states is an interesting open problem.

\subsection{Which conformal field theory?}
The decomposition $\ceff=3=2+1$ suggested by Sec.~\ref{sec:num} is not the only one. The symmetric-sector theorem shows that the theory contains the $c=\frac32$ superconformal Gaussian model of the spin-1 chain, and $3=\frac32+\frac32$ is the natural counting for a compact and a non-compact $N=1$ multiplet. In this respect it is intriguing that the phase in \eqref{eq:qII}, $[\Gamma(\frac12-\ri s)/\Gamma(\frac12+\ri s)]^2$ with $x_{\rm nc}=s^2$, coincides with the Ramond-sector reflection amplitude of $N=1$ super-Liouville theory \cite{Poghossian,RS} at its self-dual point $b=1$, and that $k=4$ is the bosonic level of the supersymmetric coset $SL(2,\mathbb{R})/U(1)$ at supersymmetric level two \cite{KS,HK}. In the latter, however, the radius of the compact direction is tied to the level, whereas here it varies with $\gamma$. A discrimination between these possibilities requires the sectors that we have not analysed: odd $S^z$, the Ising-twisted sectors such as the state with $x=X+\frac14$ above, and possible discrete states. In those sectors the roots are not all arranged in 2-strings.

\subsection{Why \texorpdfstring{$k=4$}{k=4}? Higher spin}
The $\gamma$-independence of the amplitude is already a property of the BT kernel, Eq.~\eqref{eq:WH}; what the string deviations change is its value, from $k-2=4$ to $k-2=2$, just as they change $\ceff$ from $2$ to $3$. An analytic derivation should be possible with non-linear integral equations combining the techniques of \cite{CI} with those developed for spin 1 \cite{Suzuki,HRS}, or within the ODE/IQFT correspondence \cite{BKKL1,BKKL2}; the exact relation \eqref{eq:BII} for the quasi-shift is a natural starting point.

For seas of $2S$-strings on the two lines the BT symbols generalise to
\begin{equation}
\mathsf{S}_n=\frac{2\coth G\,\sinh nG\,\sinh(P-nG)}{\sinh P},\qquad
\mathsf{A}_n=\frac{2\tanh G}{\cosh P}\times
\begin{cases}\sinh nG\,\cosh(P-nG), & n\ \text{even},\\ \cosh nG\,\sinh(P-nG), & n\ \text{odd},\end{cases}
\end{equation}
with $n=2S$, $\eta<\pi/n$ and $\hat\rho=1/(2\cosh G)$ for all $n$. Thus $\mathsf{A}_n\simeq n\eta^2\omega^2/8$ for integer spin, giving a $\gamma$-independent amplitude, whereas $\mathsf{A}_n\simeq\eta(\pi-n\eta)\omega^2/8$ for half-odd-integer spin, giving a continuously varying one. Together with $\ceff=6S/(S+1)$ this suggests a family of non-compact theories whose level is frozen for integer $S$; we have not studied the ground states for $S\ge\frac32$.

\section{Conclusions}\label{sec:concl}
We have applied the staggering construction of \cite{APSS} to higher-spin $R$-matrices and analysed the spin-1 case in detail. The local Hamiltonian \eqref{eq:H} and the quasi-shift \eqref{eq:B} have a compact form for all spins. The spin-1 model has two regimes with non-compact continuum limits: one in the universality class of the staggered six-vertex model, and a new one with $\ceff=3$, in which an exact reduction to the homogeneous spin-1 chain controls the symmetric sector, and the continuum has the black-hole form \eqref{eq:xII}--\eqref{eq:BII} at the frozen level $k=4$, with a reflection phase that does not depend on the $U(1)$ charges. The last statement rests on numerical solutions of the Bethe equations of high accuracy; its analytic derivation, the operator content beyond the 2-string sector, the fate of the exceptional states near $\gamma=\pi/4$, and the extension to $S>1$ and to open boundary conditions \cite{RJS,FG22} are left for future work.

\data{The programs used for exact diagonalization, for the solution of the Bethe equations and for the figures are provided as supplementary material.}

\ack{Author acknowledge Armenian HESC grant 24FP-1F039 for financial support.}

\appendix
\section{Exact equations for 2-strings}\label{app}
For the configuration \eqref{eq:strings}, let $\sigma_{jk}=+$ if the strings $j,k$ are on the same line and $-$ otherwise, and let $\vartheta_c(x)=2\arctan(\tan c\,\coth x)$. The phase and the modulus of Eq.~\eqref{eq:BAE} for $x_j^+$ give
\begin{align}
L\big[\Phi_{3\gamma+2\delta_j}+\Phi_{\gamma-2\delta_j}\big](\xi_j)&=2\pi I_j-\sum_{k\neq j}\sigma_{jk}\,\Theta^{\sigma_{jk}}(\xi_j-\xi_k;\delta_j,\delta_k),\\
\ln\left|\frac{\sin2\delta_j}{\sin(2\gamma+2\delta_j)}\right|&=-\frac{L}{2}\ln\frac{\cosh4\xi_j+\cos(6\gamma+4\delta_j)}{\cosh4\xi_j+\cos(2\gamma-4\delta_j)}+\frac12\sum_{k\neq j}\Lambda^{\sigma_{jk}}(\xi_j-\xi_k;\delta_j,\delta_k),
\end{align}
with
\begin{align*}
\Theta^{+}(\xi;\delta,\delta')&=\big[\theta_{\gamma+\delta-\delta'}+\theta_{\gamma-\delta+\delta'}+\theta_{2\gamma+\delta+\delta'}+\vartheta_{\delta+\delta'}\big](\xi),\\
\Theta^{-}(\xi;\delta,\delta')&=\big[\theta^\times_{\gamma+\delta-\delta'}+\theta^\times_{\gamma-\delta+\delta'}+\theta^\times_{2\gamma+\delta+\delta'}-\theta^\times_{\delta+\delta'}\big](\xi),\\
\Lambda^{\pm}(\xi;\delta,\delta')&=\ln\frac{\big[\cosh2\xi\mp\cos(2\gamma+2\delta-2\delta')\big]\big[\cosh2\xi\mp\cos(4\gamma+2\delta+2\delta')\big]}{\big[\cosh2\xi\mp\cos(2\gamma-2\delta+2\delta')\big]\big[\cosh2\xi\mp\cos(2\delta+2\delta')\big]} .
\end{align*}
The energy is $E=Ne_c-\sum_j2\,\mathrm{Re}\,\big[4\sin4\gamma/(\cosh(4\xi_j+2\ri\gamma+4\ri\delta_j)+\cos4\gamma)\big]$ and $B$ follows from \eqref{eq:Bev}. For $\delta_j\to0$ the first equation reduces to \eqref{eq:BT}.


\begin{thebibliography}{99}
\bibitem{APSS} D. Arnaudon, R. Poghossian, A. Sedrakyan, P. Sorba, Integrable chain model with additional staggered model parameter, Nucl. Phys. B 588 (2000) 638, arXiv:hep-th/0002123.
\bibitem{AASSS} J. Ambj{\o}rn, D. Arnaudon, A. Sedrakyan, T. Sedrakyan, P. Sorba, Integrable ladder $t$--$J$ model with staggered shift of the spectral parameter, J. Phys. A 34 (2001) 5887, arXiv:hep-th/0006243.

\bibitem{TSedrakyan-2001} T. Sedrakyan-Nucl.Phys.{\bf B 608} [FS] (2001) 557–576


\bibitem{ASSS} D. Arnaudon, A. Sedrakyan, T. Sedrakyan, P. Sorba, Generalization of the $U_q(gl(N))$ algebra and staggered models, Lett. Math. Phys. 58 (2001) 209, arXiv:hep-th/0106139.
\bibitem{AKMSS} D. Arnaudon, D. Karakhanyan, M. Mirumyan, A. Sedrakyan, P. Sorba, Integrable $XYZ$ model with staggered anisotropy parameter, J. Phys. A 35 (2002) 2353, arXiv:hep-th/0111233.
\bibitem{ASS} D. Arnaudon, A. Sedrakyan, T. Sedrakyan, Multi-leg integrable ladder models, Nucl. Phys. B 676 (2004) 615, arXiv:hep-th/0210087.
\bibitem{MS} V.V. Mkhitaryan, A.G. Sedrakyan, Thermodynamic Bethe ansatz for the spin-1/2 staggered $XXZ$ model, Nucl. Phys. B 673 (2003) 455, arXiv:hep-th/0302185.
\bibitem{KBS} S.A. Khachatryan, H. Babujian, A.G. Sedrakyan, Anisotropic Ising model on 2D Kagom\'e lattice as an inhomogeneous $XYZ$ integrable model, Nucl. Phys. B 1032 (2026) 117686.
\bibitem{PZ} V.Yu. Popkov, A.A. Zvyagin, ``Antichiral'' exactly solvable effectively two-dimensional quantum spin model, Phys. Lett. A 175 (1993) 295.
\bibitem{FR} H. Frahm, C. R\"odenbeck, Integrable models of coupled Heisenberg chains, Europhys. Lett. 33 (1996) 47, arXiv:cond-mat/9502090.
\bibitem{dVW92} H.J. de Vega, F. Woynarovich, New integrable quantum chains combining different kinds of spins, J. Phys. A 25 (1992) 4499.
\bibitem{JS06} J.L. Jacobsen, H. Saleur, The antiferromagnetic transition for the square-lattice Potts model, Nucl. Phys. B 743 (2006) 207, arXiv:cond-mat/0512058.
\bibitem{IJS08} Y. Ikhlef, J.L. Jacobsen, H. Saleur, A staggered six-vertex model with non-compact continuum limit, Nucl. Phys. B 789 (2008) 483, arXiv:cond-mat/0612037.
\bibitem{IJS12} Y. Ikhlef, J.L. Jacobsen, H. Saleur, An integrable spin chain for the $SL(2,\mathbb{R})/U(1)$ black hole sigma model, Phys. Rev. Lett. 108 (2012) 081601, arXiv:1109.1119.
\bibitem{Witten} E. Witten, String theory and black holes, Phys. Rev. D 44 (1991) 314.
\bibitem{DVV} R. Dijkgraaf, H. Verlinde, E. Verlinde, String propagation in a black hole geometry, Nucl. Phys. B 371 (1992) 269.
\bibitem{MOS} J. Maldacena, H. Ooguri, J. Son, Strings in AdS$_3$ and the $SL(2,\mathbb{R})$ WZW model. Part 2: Euclidean black hole, J. Math. Phys. 42 (2001) 2961, arXiv:hep-th/0005183.
\bibitem{HPT} A. Hanany, N. Prezas, J. Troost, The partition function of the two-dimensional black hole conformal field theory, JHEP 04 (2002) 014, arXiv:hep-th/0202129.
\bibitem{CI} C. Candu, Y. Ikhlef, Nonlinear integral equations for the $SL(2,\mathbb{R})/U(1)$ black hole sigma model, J. Phys. A 46 (2013) 415401, arXiv:1306.2646.
\bibitem{FS14} H. Frahm, A. Seel, The staggered six-vertex model: conformal invariance and corrections to scaling, Nucl. Phys. B 879 (2014) 382, arXiv:1311.6911.
\bibitem{BKKL1} V.V. Bazhanov, G.A. Kotousov, S.M. Koval, S.L. Lukyanov, On the scaling behaviour of the alternating spin chain, JHEP 08 (2019) 087, arXiv:1903.05033.
\bibitem{BKKL2} V.V. Bazhanov, G.A. Kotousov, S.M. Koval, S.L. Lukyanov, Scaling limit of the $\mathcal{Z}_2$ invariant inhomogeneous six-vertex model, Nucl. Phys. B 965 (2021) 115337, arXiv:2010.10613.
\bibitem{EFS} F.H.L. Essler, H. Frahm, H. Saleur, Continuum limit of the integrable $sl(2/1)$ $3$--$\bar 3$ superspin chain, Nucl. Phys. B 712 (2005) 513, arXiv:cond-mat/0501197.
\bibitem{FM12} H. Frahm, M.J. Martins, Phase diagram of an integrable alternating $U_q[sl(2|1)]$ superspin chain, Nucl. Phys. B 862 (2012) 504, arXiv:1202.4676.
\bibitem{VJS14a} E. Vernier, J.L. Jacobsen, H. Saleur, Non compact conformal field theory and the $a_2^{(2)}$ (Izergin--Korepin) model in regime III, J. Phys. A 47 (2014) 285202, arXiv:1404.4497.
\bibitem{RJS} N.F. Robertson, J.L. Jacobsen, H. Saleur, Lattice regularisation of a non-compact boundary conformal field theory, JHEP 02 (2021) 180, arXiv:2012.07757.
\bibitem{FG22} H. Frahm, S. Gehrmann, Finite size spectrum of the staggered six-vertex model with $U_q(\mathfrak{sl}(2))$-invariant boundary conditions, JHEP 01 (2022) 070, arXiv:2111.00850.
\bibitem{FGNR} H. Frahm, S. Gehrmann, R.I. Nepomechie, A.L. Retore, The $D_3^{(2)}$ spin chain and its finite-size spectrum, JHEP 11 (2023) 095, arXiv:2307.11511.
\bibitem{KL} G.A. Kotousov, S.L. Lukyanov, On the scaling behaviour of an integrable spin chain with $\mathcal{Z}_r$ symmetry, Nucl. Phys. B 993 (2023) 116269, arXiv:2305.03620.
\bibitem{ZF} A.B. Zamolodchikov, V.A. Fateev, Model factorized $S$-matrix and an integrable spin-1 Heisenberg chain, Sov. J. Nucl. Phys. 32 (1980) 298.
\bibitem{KRS} P.P. Kulish, N.Yu. Reshetikhin, E.K. Sklyanin, Yang--Baxter equation and representation theory: I, Lett. Math. Phys. 5 (1981) 393.
\bibitem{Takhtajan} L.A. Takhtajan, The picture of low-lying excitations in the isotropic Heisenberg chain of arbitrary spins, Phys. Lett. A 87 (1982) 479.
\bibitem{Babujian} H.M. Babujian, Exact solution of the isotropic Heisenberg chain with arbitrary spins: thermodynamics of the model, Nucl. Phys. B 215 (1983) 317.
\bibitem{KR} A.N. Kirillov, N.Yu. Reshetikhin, Exact solution of the integrable $XXZ$ Heisenberg model with arbitrary spin. I. The ground state and the excitation spectrum, J. Phys. A 20 (1987) 1565.
\bibitem{AGSZ} I. Affleck, D. Gepner, H.J. Schulz, T. Ziman, Critical behaviour of spin-$s$ Heisenberg antiferromagnetic chains: analytic and numerical results, J. Phys. A 22 (1989) 511.
\bibitem{AM} F.C. Alcaraz, M.J. Martins, Conformal invariance and the operator content of the $XXZ$ model with arbitrary spin, J. Phys. A 22 (1989) 1829.
\bibitem{FYF} H. Frahm, N.-C. Yu, M. Fowler, The integrable $XXZ$ Heisenberg model with arbitrary spin: construction of the Hamiltonian, the ground-state configuration and conformal properties, Nucl. Phys. B 336 (1990) 396.
\bibitem{AD} L.V. Avdeev, B.-D. D\"orfel, The Bethe-ansatz equations for the isotropic Heisenberg antiferromagnet of arbitrary spin, Nucl. Phys. B 257 (1985) 253.
\bibitem{dVW90} H.J. de Vega, F. Woynarovich, Solution of the Bethe ansatz equations with complex roots for finite size: the spin $S\ge1$ isotropic and anisotropic chains, J. Phys. A 23 (1990) 1613.
\bibitem{KB} A. Kl\"umper, M.T. Batchelor, An analytic treatment of finite-size corrections in the spin-1 antiferromagnetic $XXZ$ chain, J. Phys. A 23 (1990) L189.
\bibitem{KBP} A. Kl\"umper, M.T. Batchelor, P.A. Pearce, Central charges of the 6- and 19-vertex models with twisted boundary conditions, J. Phys. A 24 (1991) 3111.
\bibitem{Jimbo} M. Jimbo, A $q$-difference analogue of $U(\mathfrak{g})$ and the Yang--Baxter equation, Lett. Math. Phys. 10 (1985) 63.
\bibitem{BCNA} H.W.J. Bl\"ote, J.L. Cardy, M.P. Nightingale, Conformal invariance, the central charge, and universal finite-size amplitudes at criticality, Phys. Rev. Lett. 56 (1986) 742; I. Affleck, Universal term in the free energy at a critical point and the conformal anomaly, Phys. Rev. Lett. 56 (1986) 746.
\bibitem{dVW85} H.J. de Vega, F. Woynarovich, Method for calculating finite size corrections in Bethe ansatz systems: Heisenberg chain and six-vertex model, Nucl. Phys. B 251 (1985) 439.
\bibitem{Poghossian} R.H. Poghossian, Structure constants in the $N=1$ super-Liouville field theory, Nucl. Phys. B 496 (1997) 451, arXiv:hep-th/9607120.
\bibitem{RS} R.C. Rashkov, M. Stanishkov, Three-point correlation functions in $N=1$ super Liouville theory, Phys. Lett. B 380 (1996) 49, arXiv:hep-th/9602148.
\bibitem{KS} Y. Kazama, H. Suzuki, New $N=2$ superconformal field theories and superstring compactification, Nucl. Phys. B 321 (1989) 232.
\bibitem{HK} K. Hori, A. Kapustin, Duality of the fermionic 2d black hole and $N=2$ Liouville theory as mirror symmetry, JHEP 08 (2001) 045, arXiv:hep-th/0104202.
\bibitem{Suzuki} J. Suzuki, Spinons in magnetic chains of arbitrary spins at finite temperatures, J. Phys. A 32 (1999) 2341, arXiv:cond-mat/9807076.
\bibitem{HRS} \'A. Heged\H{u}s, F. Ravanini, J. Suzuki, Exact finite size spectrum in super sine-Gordon model, Nucl. Phys. B 763 (2007) 330, arXiv:hep-th/0610012.
\end{thebibliography}
\end{document}